\documentclass[11pt,a4paper]{article}
\usepackage{jheppub}
\usepackage{color}
\usepackage{mathrsfs}
\usepackage{bm}
\usepackage{amsmath}
\usepackage[stable]{footmisc}
\newcommand{\be}{\begin{equation}}
\newcommand{\ee}{\end{equation}}
\newcommand{\ba}{\begin{eqnarray}}
\newcommand{\ea}{\end{eqnarray}}
\newcommand{\nn}{\nonumber}
\newcommand{\gapp}{\mathrel{\raise.3ex\hbox{$>$}\mkern-14mu
              \lower0.6ex\hbox{$\sim$}}}
\newcommand{\lapp}{\mathrel{\raise.3ex\hbox{$<$}\mkern-14mu
              \lower0.6ex\hbox{$\sim$}}}

\newcommand{\itie}{{\it i.e.}}

\newcommand{\calD}{{\cal D}}
\newcommand{\calP}{{\cal P}}

\newcommand{\la}{\langle}
\newcommand{\ra}{\rangle}
\newcommand{\hatPhi}{\hat \Phi}

\definecolor{bittersweet}{rgb}{1.0, 0.44, 0.37}
\definecolor{coolblack}{rgb}{0.0, 0.18, 0.39}
\definecolor{britishracinggreen}{rgb}{0.0, 0.26, 0.15}
\definecolor{coolgrey}{rgb}{0.55, 0.57, 0.67}
\definecolor{darkgreen}{rgb}{0.0, 0.2, 0.13}
\definecolor{darkmagenta}{rgb}{0.55, 0.0, 0.55}
\definecolor{eggplant}{rgb}{0.38, 0.25, 0.32}
\definecolor{fashionfuchsia}{rgb}{0.96, 0.0, 0.63}

\title{Kibble mechanism for electroweak magnetic monopoles and magnetic fields}
\author{Teerthal Patel, Tanmay Vachaspati}
\affiliation{
Physics Department, Arizona State University, Tempe, AZ 85287, USA.
}
\emailAdd{tpatel28@asu.edu}
\emailAdd{tvchasp@asu.edu}

\abstract{
The vacuum manifold of the standard electroweak model is a three-sphere when one
considers  {\it homogeneous}
Higgs field configurations. For inhomogeneous configurations we argue that the vacuum 
manifold is the Hopf fibered three sphere and that this viewpoint leads 
to general criteria to detect electroweak monopoles and Z-strings.
We extend the Kibble mechanism to study the formation of electroweak monopoles and
strings during electroweak symmetry breaking. The distribution of magnetic monopoles 
produces magnetic fields that have a spectrum $B_\lambda \propto \lambda^{-2}$, where 
$\lambda$ is a smearing length scale. Even as the magnetic monopoles annihilate due to 
the confining Z-strings, the magnetic field evolves with the turbulent plasma and may be 
relevant for cosmological observations.
}


\arxivnumber{1234.5678}

\begin{document}
\maketitle
\flushbottom

\section{Introduction}
The distribution of topological defects formed 
after spontaneous symmetry breaking (SSB)
is often analyzed
by implementing the ``Kibble mechanism''~\cite{Kibble:1976sj,Vachaspati:1984dz,Ng:2008mp}. 
During SSB a field
takes on a non-trivial vacuum expectation 
value (VEV) that lies on the ``vacuum manifold''.
Distant spatial points are randomly selected on the vacuum manifold and if 
the vacuum manifold has non-trivial topology, 
the VEV of the field
may end up in a non-trivial topological configuration, in which case a topological defect would 
be formed.
Numerical simulations of the Kibble mechanism have been central to our understanding
of topological defect formation during spontaneous symmetry breaking. Notably the cosmic string 
network was shown to be dominated by infinite strings that don't close on themselves, while the 
sub-dominant distribution of closed loops was found to be scale invariant~\cite{Vachaspati:1984dz}
(for reviews see~\cite{Hindmarsh:1994re,Vilenkin:2000jqa,Kibble:2015twa}).

Here we are interested in the implications of the Kibble mechanism when the
electroweak Higgs field, denoted $\Phi$, acquires a VEV.
The electroweak vacuum manifold is a three-sphere with trivial first and second
homotopy groups and there are no 
topological magnetic monopoles or cosmic strings by these criteria. However, 
electroweak monopoles and Z-strings that connect the magnetic monopoles do exist in the model~\cite{Nambu:1977ag,Vachaspati:1992fi,Achucarro:1999it}.
We will show that a suitably modified algorithm like that in the case of topological defects can 
still be used to obtain the distribution of electroweak monopoles and strings. The distribution
can be used as an initial condition for further evolution. Since the monopoles and antimonopoles
are confined by strings, they will quickly annihilate. Yet the annihilation will leave behind a
distribution of magnetic fields~\cite{Vachaspati:1991nm,Vachaspati:1994xc}
that can be of observational interest and may have important
ramifications for 
cosmology~\cite{Durrer:2013pga,Subramanian:2015lua,Vachaspati:2020blt,Batista:2021rgm}. 

In Sec.~\ref{vacmanifold} we describe our viewpoint that the electroweak vacuum manifold
is better described as $S^2\times S^1$, {\it i.e.} as a Hopf fibered $S^3$, and thus contains
electroweak monopoles and strings.
We describe the prototype Nambu monopole in Sec.~\ref{nambu} and implement the Kibble 
mechanism in Sec.~\ref{kibble} to find a distribution of electroweak monopoles and Z-strings. 
With evolution, the network of monopoles and strings will leave behind a distribution of magnetic
fields that we characterize in Sec.~\ref{magnetic}. We summarize our conclusions in 
Sec.~\ref{conclusions}.

\section{Electroweak vacuum manifold}
\label{vacmanifold}

The vacuum manifold of the electroweak manifold is the set of all 
spatially homogeneous and static Higgs fields for which the energy function 
vanishes. The Higgs VEV is an SU(2) doublet\footnote{
For convenience we will write $\Phi$ instead of $\la \Phi \ra$ throughout this paper.
}
\be
\Phi = \begin{pmatrix} \phi_1+i\phi_2 \\ \phi_3 + i \phi_4
\end{pmatrix}
\label{Phiphi}
\ee
and since the Higgs potential is,
\be
V(\Phi ) = \lambda (|\Phi|^2-\eta^2)^2
\ee
the vacuum manifold is an $S^3$ given by
\be
|\Phi|^2 = \phi_1^2+\phi_2^2+\phi_3^2+\phi_4^2 = \eta^2.
\ee

One issue is that the symmetry of the potential consists of rotations of the 
four dimensional vector $(\phi_1,\phi_2,\phi_3,\phi_4)$, hence it is O(4), whereas
the electroweak symmetry is the smaller $[SU(2)_L\times U(1)_Y]/Z_2$. The
reduced symmetry is due to the derivative terms in the model and these are
completely ignored in discussions that are based solely on the vacuum manifold.
Derivative terms vanish for homogeneous Higgs configurations and so the
$S^3$ vacuum manifold is appropriate for such configurations.
On the other hand, the Kibble mechanism relies on VEVs that are different
in different regions of space. Hence the Higgs configurations are necessarily
inhomogeneous. We have learned from semilocal strings that the
vacuum manifold does not give the complete picture when one
considers inhomogeneous Higgs fields configuration for then the gradient
energy terms can also be important. 

Let us clarify this further by discussing the
semilocal limit of the electroweak model. Then the $SU(2)_L$ gauge coupling
is set to vanish: $g=0$. In that case, one can consider Higgs configurations that lie
entirely on the vacuum manifold but whose energy cannot vanish. This is because
the gauged $U(1)_Y$ symmetry defines $S^1$ gauge orbits on the vacuum manifold.
Only Higgs gradients along these orbits can be compensated by the gauge field 
so that the covariant gradient energy vanishes; if the Higgs VEV does not lie
on a gauge orbit, the gradient energy cannot vanish.

An alternative ``semilocal'' limit that has not previously been considered in this context
is to take the $U(1)_Y$ coupling to vanish: $g'=0$. In that case the gauge orbits
on the vacuum manifold are $S^2$'s. If we restrict attention to asymptotic Higgs 
fields configurations that have vanishing potential and gradient energy, the
Higgs VEV would have to lie on an $S^2$ and this has the right topology 
for magnetic monopoles.

The standard electroweak model has $g=0.65$ and $g'=0.34$, so neither
coupling vanishes, even though the $SU(2)_L$ coupling is larger. However
the fibered structure still exists -- the vacuum manifold $S^3$ has $S^2$ and $S^1$ 
gauge orbits. These gauge orbits are precisely defined by the Hopf fibration of $S^3$,
as was originally pointed out in the $g'=0$ semilocal limit~\cite{Gibbons:1992gt,Hindmarsh:1992ef}.
The Hopf fibration of $S^3$ provides a map from $S^3$ to $S^2$ with $S^1$ fibers.
The electroweak monopole is due to winding around the $S^2$ base manifold 
and the Z-string is due to winding around the $S^1$ fiber. Because of the
non-trivial global structure of the Hopf fibration, the Z-string is attached to the
electroweak monopole.

\section{Nambu monopole}
\label{nambu}

It is instructive to first consider the explicit configuration for the Nambu 
monopole~\cite{Nambu:1977ag} for which the asymptotic Higgs VEV is,
\be
\Phi_m = \frac{v}{\sqrt{2}} \begin{pmatrix}
	\cos(\theta/2) \\ \sin(\theta/2) e^{i\phi}
\end{pmatrix}
\label{Phim}
\ee
where $\theta$, $\phi$ are spherical angles. Note that the configuration 
is singular at $\theta=\pi$.
To see the presence of the monopole in this configuration, construct
\be
{\hat n}_m = - \hatPhi^\dag_m {\vec \sigma} \hatPhi_m
= -(\sin\theta \cos\phi, \sin\theta \sin\phi, \cos\theta ) = -{\hat r}
\label{hatnm}
\ee
where $\hatPhi \equiv \Phi /|\Phi |$, $\sigma^a$ ($a=1,2,3$) are the Pauli spin matrices
and the overall sign is chosen so that ${\hat n}={\hat z}$ when $\Phi^T=v(0,1)/\sqrt{2}$.
Now ${\hat n}_m$ is regular for all $\theta$ and $\phi$ and is in the (inner) radial direction. 
This is also called the ``hedgehog'' configuration
and immediately implies the presence of a singularity of ${\hat n}_m$ at the origin
that corresponds to a magnetic monopole~\cite{tHooft:1974kcl,Polyakov:1974ek}.
Going back to $\Phi_m$, the singularity at $\theta=\pi$ signifies the Z-string attached 
to the monopole.

The above explicit example suggests that to apply the Kibble mechanism to the electroweak 
model we should start by considering a distribution of the vector field ${\hat n}$.
Since ${\hat n}$ lives on a two-sphere ($S^2$) that has non-trivial second homotopy,
there will be hedgehog configurations of ${\hat n}$ ({\it e.g.} ${\hat n}={\hat r}$).
As for 't Hooft-Polyakov monopoles~\cite{tHooft:1974kcl,Polyakov:1974ek}, the
topological winding of ${\hat n}$ in a spherical volume of radius $R$ is given by
the surface integral
\be
n_M = \frac{1}{4\pi}
\int_R dS^i \epsilon_{abc}\epsilon_{ijk} {\hat n}^a\, \partial_j {\hat n}^b \, \partial_k {\hat n}^c .
\label{mwinding}
\ee
The discrete winding number $n_M \in \mathbb{Z}$ must remain constant as $R \to 0$. 
For $n_M\ne 0$ this implies that ${\hat n}$ is singular within the spherical volume. 

Now we consider the $\Phi$ field that corresponds to the hedgehog configuration of ${\hat n}$.
The relation between $\Phi$ and ${\hat n}$ for $|\Phi|\ne 0$ is,
\be
{\hat n} = - \hatPhi^\dag {\vec \sigma} \hatPhi 
\label{hatn}
\ee
Therefore the singularity in ${\hat n}$ for non-trivial $n_M$ requires that $\Phi=0$ at
the singular point where there is a magnetic monopole.

The Z-string attached to the monopole appears when we try and invert \eqref{hatn} to obtain 
$\Phi$. For ${\hat n}$ with non-trivial winding ($n_M\ne 0$), the reconstruction will
necessarily give a singularity in $\Phi$ (as in \eqref{Phim}). This singularity is the 
location of the Z-string on the sphere surrounding the monopole. We will describe the 
explicit algorithm for finding the location of the Z-string in Sec.~\ref{kibble}.

\section{Kibble mechanism}
\label{kibble}

The Higgs VEV of \eqref{Phiphi} can also be parametrized as,
\be
\Phi = \frac{v}{\sqrt{2}} 
\begin{pmatrix}
	\cos\alpha \, e^{i\beta} \\ \sin\alpha \, e^{i\gamma}
\end{pmatrix}
\label{PhiHopf}
\ee
where $v = 246\, {\rm GeV}$ and $\alpha \in [0,\pi/2]$, $\beta \in [0,2\pi]$, $\gamma \in [0,2\pi]$
are Hopf angular coordinates on the vacuum manifold: $\Phi^\dag \Phi = v^2/2$.  
The volume measure on the vacuum manifold in terms of Hopf coordinates is
$(1/2)d(\cos(2\alpha)) d\beta d\gamma$. Hence in any given spatial region, the values of 
$u\equiv \cos(2\alpha)$,
$\beta$ and $\gamma$ are selected from uniform probability distributions in their respective ranges.
In spatial regions that are separated by more than some
correlation length, $(u,\beta,\gamma)$ can be chosen independently. There is a lot
of theoretical and experimental literature (for a review see~\cite{Zurek:1996sj})
on the determination of the correlation length and, more recently, a full quantum
calculation for the growth of the correlation length~\cite{Mukhopadhyay:2020xmy,Mukhopadhyay:2020gwc}. 
However, the precise value of the correlation length is not a critical quantity for us since this 
only sets a length scale for the topological defects and does not
affect the scaling laws for their distribution.

In the numerical implementation we calculate the (discretized) topological winding 
for monopoles given by the surface integral in Eq.~\eqref{mwinding} as was done for
't Hooft-Polyakov monopoles~\cite{Copeland:1987ht,Leese:1990cj,Scherrer:1997sq}.
The implementation also assumes the ``geodesic rule'':  a triangular plaquette of
the spatial lattice gets mapped to a spherical triangle on the vacuum manifold, but three
points on a two-sphere define two complementary spherical triangles and we choose the 
one with the smaller area~\cite{Vachaspati:1984dz,Pogosian:1997ez}.

\begin{figure}
	\includegraphics[width=0.5\textwidth,angle=0]{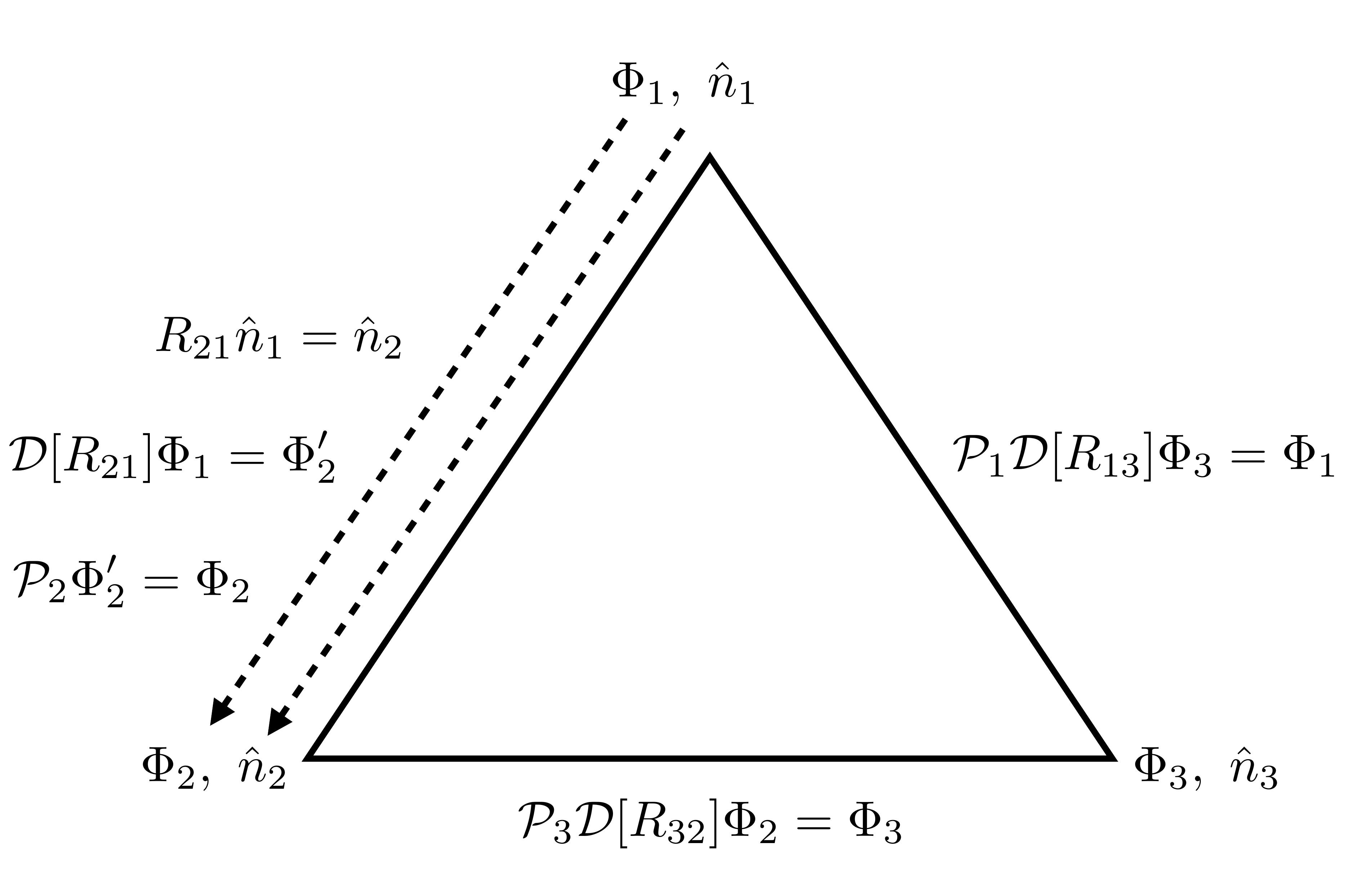}
	\centering
	\caption{A triangular plaquette is assigned values of $\Phi$ at its vertices, from which
		we determine corresponding values of ${\hat n}$ using \eqref{hatn}. We find the
		rotation $R_{21}$ that takes ${\hat n}^1$ to ${\hat n}^2$. This rotation in $SO(3)$
		also defines a rotation, ${\cal D}[ R_{21}]$ in $SU(2)_L$ that acts on $\Phi_1$ to
		give $\Phi_2'$ which in general differs from $\Phi_2$ by rotation by an element 
		${\cal P}_2 \in U(1)_Z$, at vertex 2. Similarly we can obtain the
		rotations that take $\Phi_2$ to $\Phi_3$, and $\Phi_3$ to $\Phi_1$. The total rotation
		in going from vertex 1 around the triangle and back to vertex 1 is:
		${\cal P}_1 {\cal D}[ R_{13}] {\cal P}_3 {\cal D}[ R_{32}] {\cal P}_2 {\cal D}[ R_{21}] $,
		and this rotation acts on $\Phi_1$ to give back $\Phi_1$. If the net Z-phase rotation in 
		going around the plaquette is $\pm 2\pi$, there
		is a Z-string (or anti-string) passing through the plaquette. 
	}
	\label{stringscheme}
\end{figure}

We now turn to the Z-strings that connect the monopoles.
First we note that ${\hat n}$ is invariant under $K \equiv [U(1)_L\times U(1)_Y]/Z_2$ 
transformations, where $U(1)_L \subset SU(2)_L$ consists of rotations about the axis
${\hat n}$ and $U(1)_Y$ are phase rotations of $\Phi$. (The $Z_2$ consists of the common
elements, $\pm {\bf 1}$, contained in both $U(1)_L$ and $U(1)_Y$.) 
The group $K$
can also be thought of as $U(1)_Z \times U(1)_Q$ where $Q$ denotes the generator
of the electromagnetic group and is given by
$Q = ({\bf 1} + {\hat n}\cdot {\vec \sigma} )/2$.
The generator of $U(1)_Z$ is
\be
T_Z = \frac{{\bf 1} - {\hat n}\cdot {\vec \sigma} }{2}.
\label{TZ}
\ee
The VEV of $\Phi$ is invariant under the electromagnetic $U(1)_Q$ since $Q\Phi=0$.
Thus, for a fixed ${\hat n}$, there is an entire circles worth of $\Phi$'s given by rotations by $U(1)_Z$. 
As we go around a spatial plaquette, rotations of the ${\hat n}$ vectors define ``parallel transport'' 
of the $\Phi$ fields, which may differ from the actual $\Phi$ by an element of $U(1)_Z$, as explained in
Fig.~\ref{stringscheme}. Non-trivial winding of the $U(1)_Z$ phase factor implies the existence
of a Z-string passing through the plaquette.

Consider one leg of a triangular plaquette as shown in Fig.~\ref{stringscheme}. The
vector ${\hat n}_1$ is rotated into ${\hat n}_2$, \itie~${\hat n}_2 = R_{21}{\hat n}_1$, by an 
$SO(3)$ rotation about the axis ${\hat a}_{21}$ and by angle $\theta_{21}$,
\be
{\hat a}_{21} = \frac{{\hat n}_1\times {\hat n}_2}{|{\hat n}_1\times {\hat n}_2|}, \ \ 
\theta_{21} = \cos^{-1} ({\hat n}_1 \cdot {\hat n}_2 )
\ee
and we take $0 \le \theta_{12} \le \pi$.
A corresponding $SU(2)_L$ rotation is\footnote{There are two elements of $SU(2)_L$,
	namely $\pm {\cal D}[R_{21}]$, that correspond to the $SO(3)$ rotation $R_{21}$. This
	ambiguity will be absorbed in ${\cal P}_2$ defined in \eqref{P2def} as ${\cal P}_2$
	also gives a phase factor in its action on $\Phi_2'$ as shown in \eqref{TZ2onPhi2'}.}
\be
\calD[R_{21} ]= \exp\left ( -i {\hat a}_{21}\cdot {\vec \sigma} \frac{\theta_{21}}{2} \right )
\label{calD}
\ee
and rotates $\Phi_1$ to,
\be
\Phi_2' = \calD[R_{21}] \Phi_1
\ee
In general, $\Phi_2' \ne \Phi_2$ and an additional $U(1)_Z$ rotation, $\calP_2$, may be necessary
to rotate $\Phi_1$ to $\Phi_2$,
\be
\Phi_2 = \calP_2 \, \Phi_2' = \calP_2 \, \calD [R_{21}] \Phi_1
\label{P2def}
\ee
where $\calP_2 = e^{i T_{Z2} \delta_2}$. $T_{Z2}$ is as defined in \eqref{TZ} with ${\hat n}={\hat n}_2$,
and $\delta_2$ is a phase angle. To determine $\delta_2$ we use,
\be
e^{i\delta_2} = \Phi_2'^\dag \Phi_2 
\ee
which can be derived using \eqref{calD}. We will choose $\delta_2$ with the smallest value 
of $|\delta_2|$ in accordance with the geodesic rule~\cite{Vachaspati:1984dz,Pogosian:1997ez}. 
Note that $\calP_2 = e^{i T_{Z2} \delta_2}$ acts on $\Phi_2'$ to simply give a phase 
factor $\exp(i\delta_2)$,
\be
e^{i T_{Z2} \delta_2} \Phi_2' = e^{i\delta_2} \Phi_2'
\label{TZ2onPhi2'}
\ee
because ${\hat n}_2 = - \hatPhi_2^\dag {\vec \sigma} \hatPhi_2 = - \hatPhi_2'^\dag {\vec \sigma} \hatPhi_2'$.

In this way we can go around all the sides of the triangular plaquette and obtain
\be
\Phi_1 = \calP_1 \, \calD [R_{13}] \calP_3 \, \calD [R_{32}] \calP_2 \, \calD [R_{21}] \Phi_1
\equiv {\bf R} \Phi_1
\label{Phi1Phi1}
\ee
The right-most rotation, $\calD [R_{21}] \Phi_1$, yields
$\Phi_2'$ and, as in \eqref{TZ2onPhi2'}, the action of $\calP_2$ acting on $\Phi_2'$ simply
gives a phase factor that commutes with all other rotations in \eqref{Phi1Phi1}. Hence the
action of $\calP_2$ is to give an overall factor of $e^{i\delta_2}$. Similar arguments apply
to the action of $\calP_1$ and $\calP_3$. Then the action of ${\bf R}$ on $\Phi_1$ is
equivalent to multiplication by,
\be
{\bf R}= e^{i (\delta_1+\delta_2+\delta_3 + h_{123})}
\ee
where $h_{123}$ denotes the phase angle due to the rotation 
$\calD [R_{13}] \calD [R_{32}] \calD [R_{21}]$. This rotation implements
the parallel transport of $\Phi_1$ all the way around the triangular plaquette 
and gives the holonomy angle, $h_{123}$, in this process. To determine $h_{123}$
we use
\be
e^{ih_{123}}= 
\Phi_1^\dag \calD [R_{13}] \calD [R_{32}] \calD [R_{21}] \Phi_1
\ee
From \eqref{Phi1Phi1} we must have
\be
\delta_1+\delta_2+\delta_3 + h_{123} = 0, \pm 2\pi
\label{swinding}
\ee
and a value of $\pm 2\pi$ signals that a Z-string/anti-string passes through the plaquette.

\begin{figure}
      \centering
      \includegraphics[width=0.35\textwidth,angle=0]{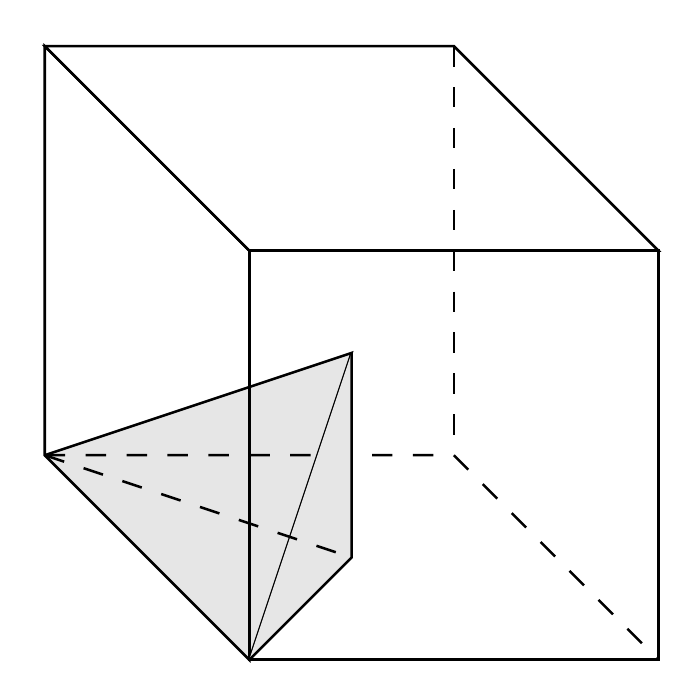}
        \caption{The cubic lattice is divided into tetrahedral cells in our simulations.}
  \label{tetrahedralCell}
\end{figure}

\begin{figure}
	\centering
	\includegraphics[width=0.42\textwidth,angle=0]{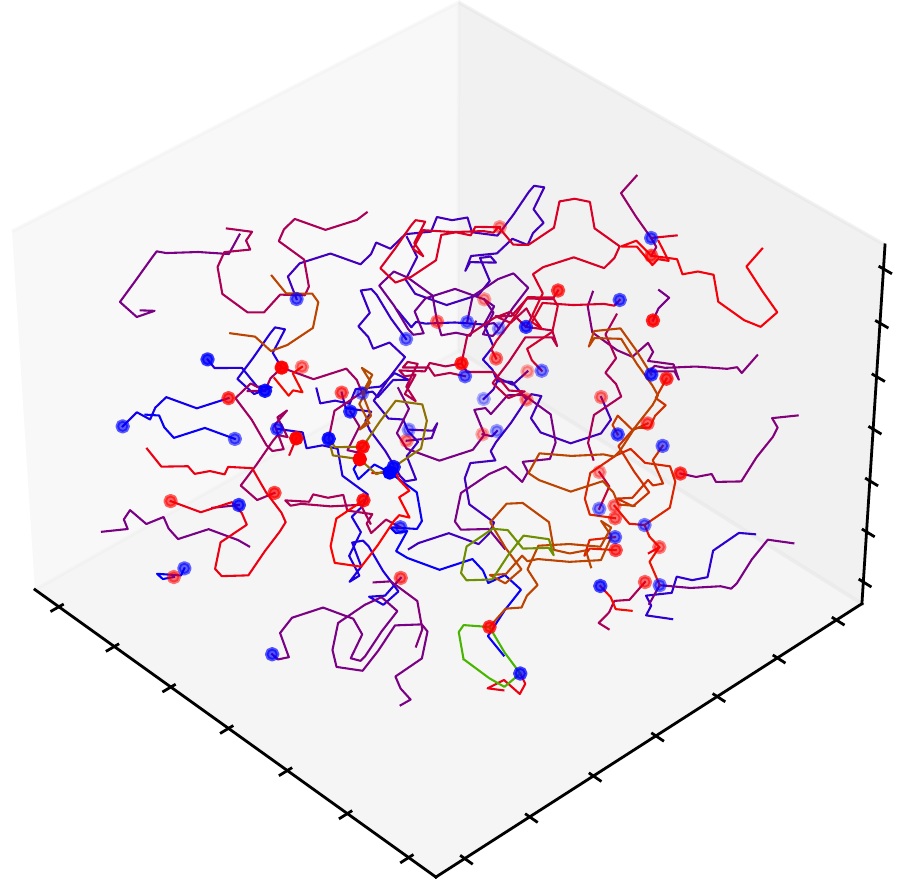}
	\caption{Sample monopole distribution with strings connecting them. Some of the
		strings are in the form of closed loops.}
	\label{picture}
\end{figure}

\begin{figure}
	\includegraphics[width=0.5\textwidth,angle=0]{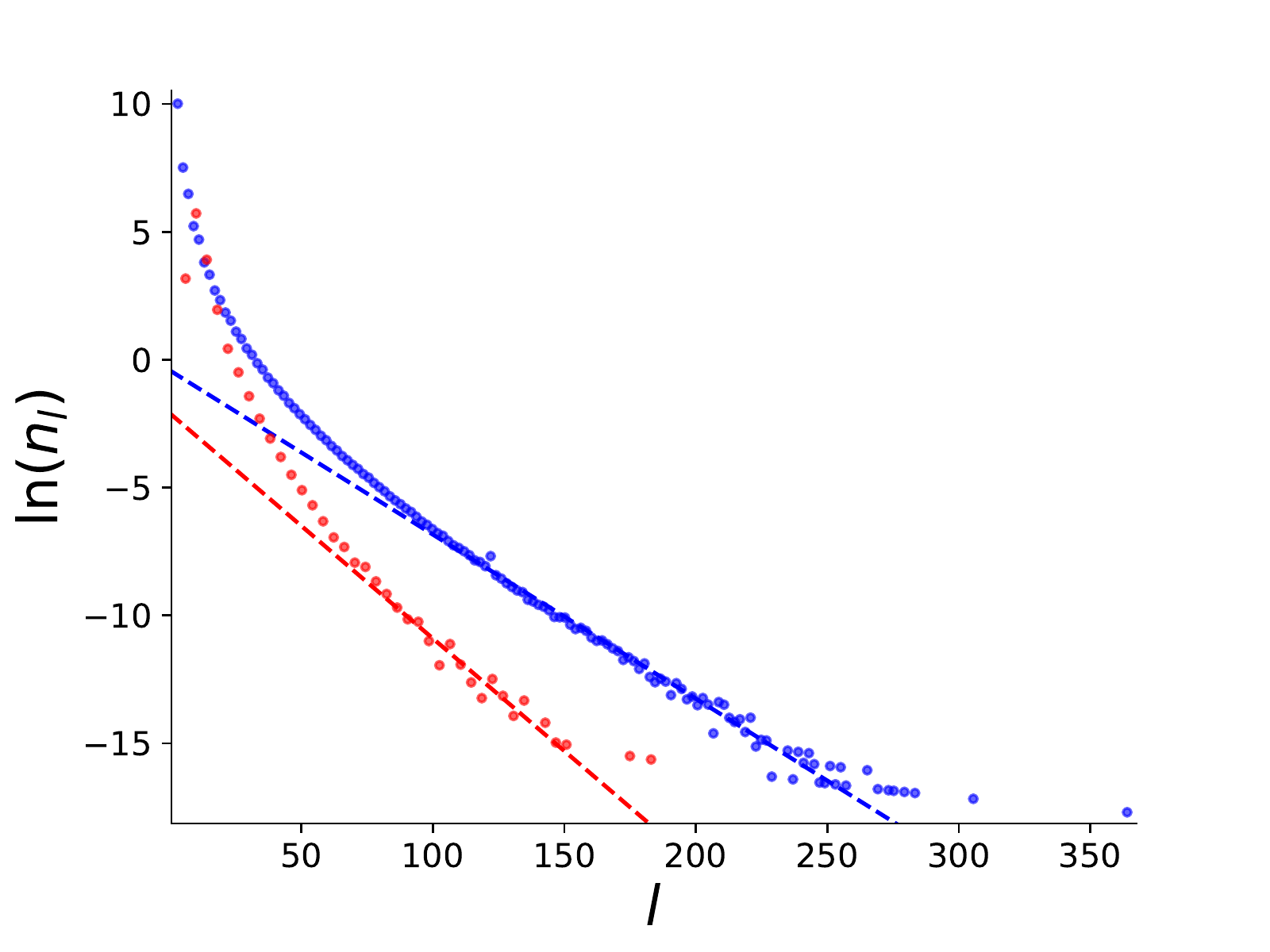}
	\centering
	\caption{Log-linear plot of number density of open strings (blue) and closed strings (red)
		vs. length $l$. The parameters of the dashed fitting curves are given in \eqref{dnopen}
		and \eqref{dnclosed}.
	}
	\label{nlvsl}
\end{figure}

We have numerically implemented this algorithm to study the distribution of monopoles
and strings on a discrete tetrahedral lattice. Each cell of a cubic lattice is divided into
24 tetrahedra~\cite{Ng:2008mp} as shown in Fig.~\ref{tetrahedralCell}.
At every lattice point, we assign random values of $\alpha$, $\beta$
and $\gamma$, from which we construct $\Phi$ and ${\hat n}$. We find the monopoles on 
the lattice by evaluating the 
monopole winding in \eqref{mwinding} for every tetrahedral cell, and the strings are found
by evaluating the winding in \eqref{swinding} for every triangular plaquette.
A sample of the monopole distribution with strings is shown in 
Fig.~\ref{picture}. 

As in earlier simulations of monopole formation~\cite{Copeland:1987ht,Leese:1990cj,Scherrer:1997sq},
${\hat n}$ is uniformly distributed on an $S^2$ and the magnetic
charge within a volume, $\sim L^3$, is given by a surface integral due to Gauss' law, with $N \sim (L/\xi)^2$ 
independent domains of size $\xi$ on the surface. Hence the root-mean-square magnetic charge 
within the volume goes as $\sqrt{N} \sim L/\xi$. We have confirmed this scaling in our simulations.

We also evaluate the length distribution of open string segments, {\it i.e.} the number density of strings of
length between $l$ and $l+dl$, denoted $dn_{\rm open}(l)$. The dependence of $dn_{\rm open}(l)$ on $l$ 
is shown in Fig.~\ref{nlvsl} and is fit by a decaying exponential,
\ba
&&dn_{\rm open} (l) = A_o e^{-l/l_o}\, dl, \nn \\
&&A_o = 0.12 \pm 0.06 ,\ \  l_o = 6.68\pm 0.28
\label{dnopen}
\ea
where the length is measured in units of the step length in going from one
tetrahedral cell to its neighboring cell.
The number density of closed loops also follows an exponential with,
\be
A_c = 0.66\pm 0.07,\ \  l_c =  7.79\pm 0.08.
\label{dnclosed}
\ee

\section{Magnetic field}
\label{magnetic}

As in the case of topological defects, the Kibble mechanism only provides initial conditions
for the evolution of the system. In the case of cosmic strings, small loops formed during
the symmetry breaking will quickly 
collapse and dissipate, while longer loops and infinite strings will persist and eventually
reach a scaling solution. In the electroweak case, monopoles and anti-monopoles will be
brought together by the confining strings and rapidly annihilate~\cite{Everett:1984yc}. 
However their annihilation will leave behind a magnetic
field. Since Maxwell equations hold after electroweak symmetry breaking, the
magnetic field can then be evolved with the usual Maxwellian magneto-hydrodynamical (MHD)
equations~\cite{Brandenburg:2017neh}. We now turn to a characterization of the initial magnetic 
field.

The electromagnetic field strength is defined as
\be
A_{\mu\nu} = \partial_\mu A_\nu - \partial_\nu A_\mu
 - i \frac{2\sin\theta_w}{g} (\partial_\mu\hatPhi^\dag \partial_\nu \hatPhi -
\partial_\nu \hatPhi^\dag \partial_\mu\hatPhi )
\label{Adef}
\ee
where $A_\mu \equiv \sin\theta_w {\hat n}^a W_\mu^a + \cos\theta_w Y_\mu$
and the last term in \eqref{Adef} is required for a suitable gauge invariant definition
of $A_{\mu\nu}$~\cite{tHooft:1974kcl,Vachaspati:1991nm}. The definition breaks
down at points where $|\Phi |=0$, {\it i.e.} in the symmetry restored phase, because
${\hat n}$ and $\hatPhi$ are not well-defined.

The magnetic field of the monopole is
\be
{\bf B} = \nabla \times {\bf A} 
- i \frac{2\sin\theta_w}{g} \nabla \hatPhi^\dag \times \nabla \hatPhi
\label{BPhi}
\ee
With $\Phi=\Phi_m$ of Eq.~\eqref{Phim} and ${\bf A}=0$ we find the monopole
magnetic field outside the core of the monopole,
${\bf B}_m = \sin\theta_w \hat r /(gr^2)$
where $r$ is the radial coordinate.
Around the Z-string at $\theta=\pi$ we find $\hatPhi_m \to e^{i\phi} (0,1)^T$. 
Using this form in \eqref{BPhi} we see that there is no electromagnetic field
associated with the Z-string at locations where $\Phi \ne 0$.
We can extend the formula \eqref{BPhi} to the point where $\Phi=0$ in the Z-string 
by using continuity, and then the magnetic field vanishes everywhere for the Z-string.

The usual characterization of stochastic isotropic magnetic fields is in terms of
the two point correlators,
\be
\langle B_i ({\bf x}+{\bf r}) B_j ({\bf x}) \rangle = 
M_N(r) (\delta_{ij}-{\hat r}_i{\hat r}_j)+M_L(r)  {\hat r}_i {\hat r}_j
+ \epsilon_{ijk} r_k M_H(r)
\label{gencor}
\ee
In Maxwell theory, the correlation functions $M_N$ and $M_L$ are related by 
the condition that the magnetic field is divergence free,
\be
\frac{1}{2r} \frac{d}{dr} \left ( r^2 M_L(r) \right ) = M_N(r).
\ee
In our case, however, the magnetic field is 
not divergence-free and $M_N$ and $M_L$ are independent functions. The helical
correlator, $M_H$, vanishes for us since we have not included any source of parity 
violation in the system. 

We have evaluated the magnetic field correlator numerically and find
\be
\langle B_i ({\bf x}+{\bf r}) B_j ({\bf x}) \rangle  = f(r) \delta_{ij}
\label{cor}
\ee
with $f(r)$ exhibiting anti-correlations at small scales. This makes physical
sense since it is known that defects are preferentially surrounded by 
anti-defects~\cite{Leese:1990cj}.

Once the monopoles and antimonopoles have annihilated, the correlator in \eqref{cor}
should revert to the form in \eqref{gencor} with the standard divergence free condition.
We have not yet studied this evolution. Instead we use a ``smearing procedure''
to estimate the volume averaged magnetic field due to monopoles,
\be
\la {\bf B} \ra_V = \frac{1}{V} \int_V d^3x\, {\bf B}
= -i \frac{2\sin\theta_w}{g V} \int_{\partial V} d{\bf S} \times (\hatPhi^\dag \nabla\hatPhi)
\label{volavg}
\ee 
where the last expression for the surface integral follows from using \eqref{BPhi} together 
with an integration by parts. Note that \eqref{BPhi} assumes $|\Phi| \ne 0$ and hence
is not valid in the interior of the integration volume $V$ in the presence of monopoles. 
The volume integral in \eqref{volavg} is ambiguous because of the divergent magnetic field 
at the locations of the monopoles. However the surface integral given in \eqref{volavg} 
still applies as the surface of integration does not intersect any monopole cores. 
The surface may intersect Z-strings but the formula in \eqref{BPhi} 
holds by continuity as discussed below \eqref{BPhi}.

\begin{figure}
	\includegraphics[width=0.50\textwidth,angle=0]{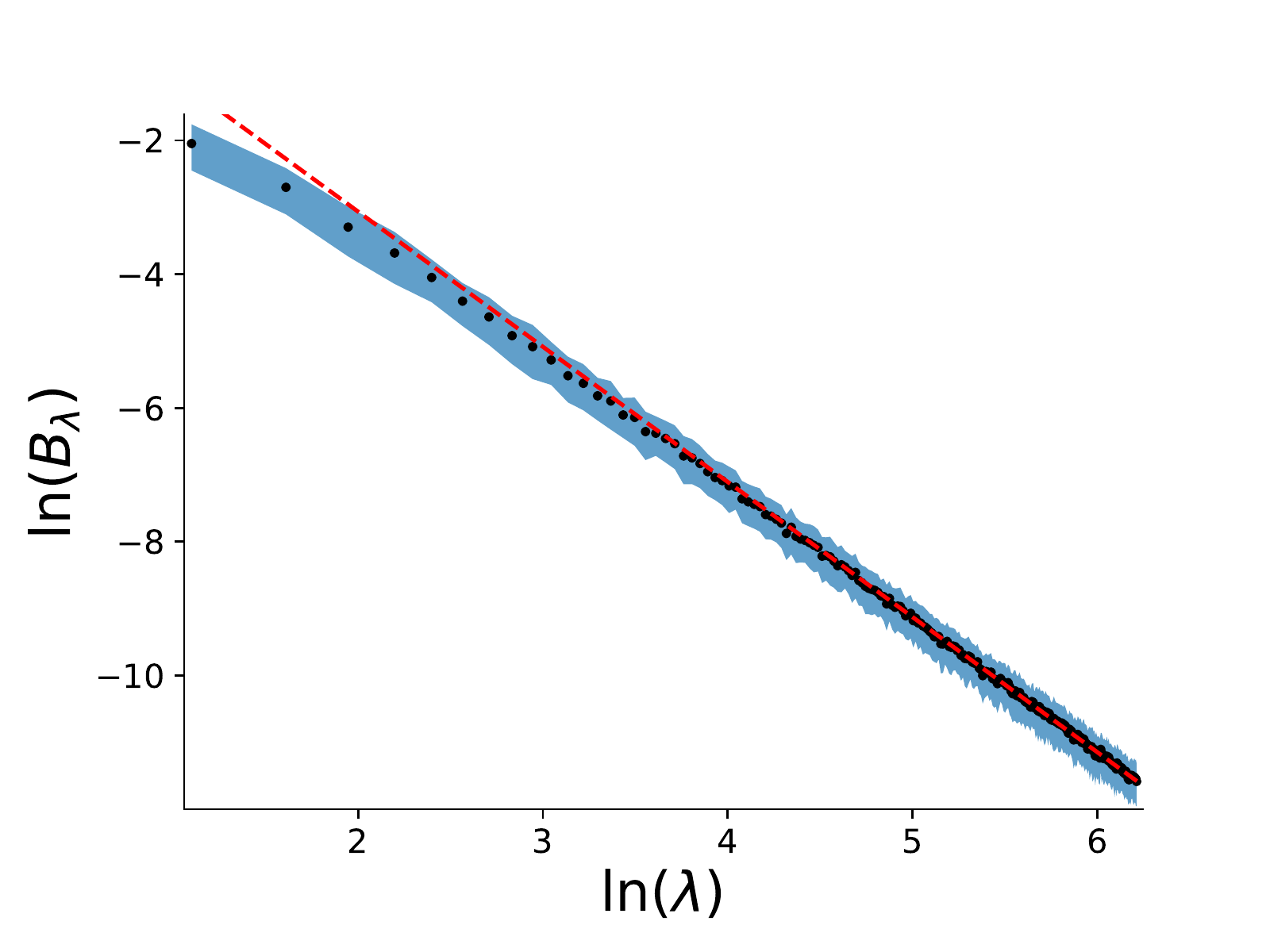}
	\centering
	\caption{Log-log plot of the smeared magnetic field strength, $B_\lambda$, vs.
		$\lambda$. The blue band shows the 1-$\sigma$ spread of the individual Monte Carlo 
		results. The dashed line shows the fit 
		$\ln(B_\lambda)=(-2.02\pm 0.02)\ln(\lambda) + (0.98\pm 0.09)$.}
	\label{Blambdafig}
\end{figure}

For the integration in \eqref{volavg} we will consider cubical volumes with side $\lambda$. 
If $\xi$ denotes the size of domains in which the random variable
$\hatPhi^\dag \nabla \hatPhi$ is tightly correlated, the discretized surface integral
in \eqref{volavg} consists of a sum of $(\lambda/\xi )^2$ independent random terms
and the sum itself will go like the square root of this number.
Therefore we expect the magnitude $B_\lambda \equiv | \la {\bf B} \ra_V |$
to grow as $B_\lambda \propto \lambda/V \propto 1/\lambda^2$.
We have numerically evaluated $B_\lambda$ and the result is plotted in 
Fig.~\ref{Blambdafig}. The fit shows indeed shows that $B_\lambda \propto 1/\lambda^2$.

As a final comment, note that the numerical calculation of the magnetic field does not
directly use the network of monopoles and strings discussed in the previous sections. 
All that is needed is to evaluate the final term of \eqref{BPhi} from the random distribution 
of the Higgs VEV.

\section{Conclusions}
\label{conclusions}

Vacuum configurations of a field theory should include all configurations with
minimum energy. Conventional considerations focus on homogeneous fields
and then the vacuum manifold is given by the minima of the potential. However,
in gauge theories, inhomogeneous configurations can also have minimum
energy provided they lie on gauge orbits on the vacuum manifold. Thus the 
vacuum manifold has additional structure. In particular, by minimizing the
potential of the electroweak model the vacuum manifold is seen to be an $S^3$.
However the gauge orbits map the $S^3$ to $S^2$ with $S^1$ fibers, {\it i.e.} the
vacuum manifold is a Hopf fibered $S^3$. The topology of $S^2 \times S^1$
leads to electroweak magnetic monopoles that are confined by Z-strings whose
distribution we have determined by an extension of the Kibble mechanism.
Since the electroweak monopoles are confined by Z-strings, they will annihilate
rapidly even as they are formed, leaving behind a cosmological magnetic field
whose spectrum falls off slowly with increasing wavelength: $B_k \propto k^2$.

An alternative approach to deriving the properties of the magnetic field is to directly
simulate the electroweak symmetry breaking, as has been done in several 
works~\cite{DiazGil:2007dy,DiazGil:2008tf,Ng:2010mt,Mou:2017zwe,Zhang:2019vsb}.
These field theory simulations are much more computationally intensive than the present
approach and are limited by computer resources. On the flip side, an advantage is that
they more completely account for the dynamical evolution during the symmetry breaking, 
including magnetic fields that may be generated independently of the monopoles (the 
$A_\mu$ terms in \eqref{Adef}). 

The MHD evolution of magnetic fields depends significantly on the helicity of the field, 
described by the parity odd $M_H$ correlator in \eqref{gencor}. There is, however, no
source of parity violation in the formulation of the Kibble mechanism, and indeed in the 
bosonic sector of the electroweak model. Hence the magnetic field will be (globally) 
non-helical. (The process of monopole annihilation can induce local helicity
because, in general, the monopole and antimonopole will be relatively
twisted~\cite{Vachaspati:2015ahr}.)
It is an interesting open question if parity violation from the fermionic sector or
extensions of the standard model can be incorporated in the Kibble mechanism,
that can then be used to study the generation of helical magnetic fields. Parity violating
effects are also necessary for generating cosmic matter-antimatter asymmetry and 
the connection with magnetic helicity has already been 
noted~\cite{Vachaspati:1994ng,Cornwall:1997ms,Vachaspati:2001nb,Copi:2008he,
	Chu:2011tx,Jackiw:1999bd,Zhang:2017plw}.

The evolution of the magnetic field from the electroweak epoch to the present epoch
is affected by several factors: turbulence, cosmic expansion, dissipation, and
perhaps novel chiral effects. Magneto-hydrodynamical evolution does not apply initially 
because the magnetic field is not divergence-free.
From general arguments that are supported by numerical simulations,
a few percent of the electroweak false vacuum energy goes into magnetic fields during 
spontaneous symmetry breaking~\cite{Vachaspati:2020blt}. The coherence scale of the 
magnetic field at the electroweak epoch, $\xi (t_{EW})$, will depend on the dynamics during 
electroweak symmetry breaking. To obtain estimates we use an upper bound on the
coherence and take it to be the horizon size at the electroweak scale: $\xi(t_{EW}) \sim
t_{EW} \sim 1~{\rm cm}$. Then the magnetic field on length scale $\lambda$ at the
present epoch is given by
\be
B_\lambda (t_0) \sim \sqrt{\rho_\gamma (t_0)} \left ( \frac{t_{EW}}{\lambda (t_{EW}) } \right )^2
\ee
where $\lambda(t_{EW}) = \lambda(t_0) T_{0}/T_{EW}$ and $T$ denotes the 
cosmic temperature. With $T_0 \sim 10^{-4}~{\rm eV}$, $T_{EW} \sim 10^{11}~{\rm eV}$,
$\sqrt{\rho_\gamma (t_0)} \sim 10^{-6}~{\rm G}$, we get
\be
B_{1\,{\rm kpc}} (t_0) \sim 10^{-18}~{\rm G}.
\ee
and on Mpc scales the magnetic field is $\sim 10^{-24}~{\rm G}$. 
This estimate is much smaller than the blazar lower bounds in the literature: 
$B_{1\, {\rm Mpc}} \gtrsim 10^{-16}-10^{-19}~{\rm G}$
~\cite{Neronov73,Essey_2011,Finke_2015,Biteau:2018tmv}.
Hence the monopoles by themselves cannot provide
magnetic fields of the observed strength. Additional ingredients are necessary
if the magnetic fields generated during electroweak symmetry breaking are to
explain observations. In particular, magnetic helicity can be this necessary ingredient
as it can stretch the coherence scale of the magnetic field by a large factor $\sim 10^7$
and increase the field strength estimate to $\sim 10^{-11}\, {\rm G}$ on 10~kpc 
scales~\cite{Vachaspati:2020blt}.

In summary, we have extended the Kibble mechanism and applied it to the electroweak
model. Then topological considerations 
lead to a distribution of magnetic monopoles and Z-strings that we can characterize. The 
distribution of magnetic monopoles immediately implies the presence of magnetic fields. We 
have derived the (smeared) magnetic field distribution as a function of the smearing length
scale, $\lambda$, and find $B_\lambda \propto \lambda^{-2}$. The role of early universe
magnetic fields for cosmological observations has been recently reviewed
in Refs.~\cite{Durrer:2013pga,Subramanian:2015lua,Vachaspati:2020blt,Batista:2021rgm}.

\section{Acknowledgments}

We are grateful to Heling Deng, Alan Guth, Ken Olum and Alex Vilenkin for comments and 
to  Heling Deng for numerical help.
This work was supported by the U.S. Department of Energy, Office of High Energy Physics, 
under Award DE-SC0019470 at ASU.

\newpage

\bibliographystyle{jhep}
\bibliography{paper}

\end{document}